\documentclass[11pt,a4paper]{article}
\usepackage{jheppub}
\usepackage{tikz}
\usepackage{amsmath}
\usepackage{graphicx}
\usepackage{dcolumn}
\usepackage{xspace}
\usepackage{color}
\usepackage{url}
\usepackage[utf8]{inputenc}
\usepackage{epstopdf}

\newcommand{\bea}{\begin{eqnarray}}
\newcommand{\eea}{\end{eqnarray}}
\newcommand{\nn}{\nonumber\\}

\newcommand{\Br}{\mathrm{Br}}

\newcommand{\I}{\mathrm{i}}
\DeclareMathOperator{\sgn}{sgn}

\usetikzlibrary{decorations.pathmorphing}
\usetikzlibrary{decorations.pathreplacing}
\usetikzlibrary{decorations.markings}
\usetikzlibrary{arrows}

\tikzset{
	photon/.style={decorate, decoration={snake}},
	fermion/.style={postaction={decorate},
		decoration={markings,mark=at position .55 with {\arrow{>}}}},
	antifermion/.style={postaction={decorate},
		decoration={markings,mark=at position .55 with {\arrow{<}}}}
}

\definecolor{math1}{rgb}{0.368417,0.506779,0.709798}
\definecolor{math2}{rgb}{0.880722,0.611041,0.142051}
\definecolor{math3}{rgb}{0.560181,0.691569,0.194885}
\definecolor{math4}{rgb}{0.922526,0.385626,0.209179}
\definecolor{math5}{rgb}{0.528488,0.470624,0.701351}
\definecolor{math6}{rgb}{0.772079,0.431554,0.102387}
\definecolor{math7}{rgb}{0.363898,0.618501,0.782349}
\definecolor{math8}{rgb}{1,0.75,0}
\definecolor{red2}{rgb}{1,0.8,0.8}
\definecolor{red4}{rgb}{1,0.6,0.6}

\definecolor{math1D}{rgb}{0.245611,0.337853,0.473199}
\definecolor{math1DD}{rgb}{0.163741,0.225235,0.315466}

\newcommand{\legline}[4]{
	\draw[color=#1,#4] (#2) --+ (0.6,0) node [right,xshift=5] {\color{black}#3};
}
\newcommand{\legbox}[3]{
	\draw[color=#1,fill] (#2) --+ (-0.1,0) --+ (-0.1,0.1) --+ (0.1,0.1) --+ (0.1,-0.1) --+ (-0.1,-0.1) --+ (-0.1,0) node [right,xshift=10] {\color{black}#3};
}

\begin{document}
\thispagestyle{empty}
\begin{flushright}
PSI-PR-17-11\\
ZU-TH 16/17\\
\today\\
\end{flushright}
\vspace{3em}
\begin{center}
{\Large\bf Correlating Lepton Flavour (Universality) Violation in $B$ Decays with $\mu\to e\gamma$ using Leptoquarks}
\\
\vspace{3em}
{\sc Andreas Crivellin$^a$, Dario M\"uller$^{a,b}$, A.~Signer$^{a,b}$, Y. Ulrich$^{a,b}$
}\\[2em]
{\sl ${}^a$ Paul Scherrer Institut,\\
CH-5232 Villigen PSI, Switzerland \\
\vspace{0.3cm}
${}^b$ Physik-Institut, Universit\"at Z\"urich, \\
Winterthurerstrasse 190,
CH-8057 Z\"urich, Switzerland}
\setcounter{footnote}{0}
\end{center}
\vspace{2ex}
\begin{center}
\begin{minipage}[]{0.9\textwidth}
{} {\sc Abstract:}\\Motivated by the measurements of $b\to s\ell^+\ell^-$
		transitions, including $R(K)$ and $R(K^*)$, we examine lepton flavour
		(universality) violation in $B$ decays and its connections to $\mu\to
		e\gamma$ in generic leptoquark models. Considering all 10
		representations of scalar and vector leptoquarks under the Standard
		Model gauge group we compute the tree-level matching for semileptonic
		$b$-quark operators as well as their loop effects in
		$\ell\to\ell^\prime\gamma$.  In our phenomenological analysis, we
		correlate $R(K)$, $R(K^*)$ and the other $b\to s\mu^+\mu^-$ data to
		$\mu\to e\gamma$ and $b\to s\mu e$ transitions for the three
		leptoquark representations that generate left-handed currents in
		$b\to s\ell^+\ell^-$ transitions and, therefore, provide a good fit to
		data.  We find that while new physics contributions to muons are
		required by the global fit, also couplings to electrons can be
		sizeable without violating the stringent bounds from $\mu\to e\gamma$.
		In fact, if the effect in electrons in $b\to s\ell^+\ell^-$ has the
		opposite sign from the effect in muons, the bound from $\mu\to e\gamma$
		can always be avoided.  However, unavoidable effects in $b\to s\mu e$
		transitions (i.e.  $B_s\to\mu e$, $B\to K\mu e$, etc.) appear that
		are within the reach of LHCb and BELLE II.
\end{minipage}
\end{center}

\setcounter{page}{1}


\bigskip

	\section{Introduction}
	\label{intro}
	
	The LHC completed the Standard Model (SM) of particle physics by
	discovering the Higgs boson but it did not yet directly observe any
	particles beyond the ones already present in the SM. However, several
	measurements of $b\to s\mu^+\mu^-$ transitions in recent years have
	lead to a tension with SM predictions. Due to an intriguing pattern in
	these anomalies it is tempting to interpret them as an indirect hint
	for new physics (NP)~\cite{Altmannshofer:2015sma,
		Descotes-Genon:2015uva, Hurth:2016fbr}. Taking this approach and
	including the new LHCb result~\cite{Aaij:2017vbb} for $R(K^*)=(B\to
	K^*\mu^+\mu^-)/(B\to K^*e^+e^-)$, measuring lepton flavour universality (LFU)
	violation, the global significance for NP increased above the
	$5\,\sigma$ level~\cite{Capdevila:2017bsm}. In addition, the
	combination of the ratios $R(D^{(*)})=(B\to D^{(*)} \tau\nu)/(B\to
	D^{(*)} \ell\nu)$ also differs by $3.9\,\sigma$ from its SM
	prediction~\cite{Amhis:2016xyh}.  All together, this strongly
	motivates us to examine LFU violation in semileptonic $B$ decays in the
	context of NP.
	
		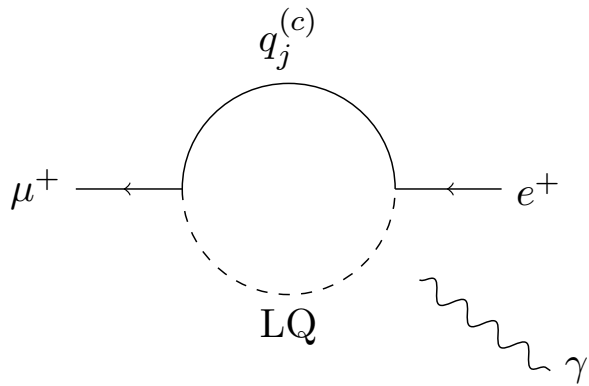
\begin{figure*}[t]
		\centering
\scalebox{1.4}{
		\begin{tikzpicture}
		\draw[fermion] (-1,0) -- (-2,0) node[left]{$\mu^+$};
		\draw[antifermion] ( 1,0) -- ( 2,0) node[right]{$e^+$};
		\draw (-1,0) arc[radius =1, start angle=180, end angle=0] node [midway,above]{$q^{(c)}_j$};
		\draw[dashed] (-1,0) arc[radius =-1, start angle=0, end angle=180] node [midway,below]{LQ};
		\draw[photon] (-35:1.5) --+ (-35:1.5) node [right] {$\gamma$};
		\end{tikzpicture}
}
		\caption{Feynman diagram generating $\mu\to e\gamma$ in models with
			leptoquarks.}
		\label{FeynmanDiagrams}
	\end{figure*}
	
	Since $b\to s\ell^+\ell^-$ processes are semileptonic, leptoquarks
	(LQ) provide a natural explanation for these anomalies (see, for
	example,~\cite{Gripaios:2014tna,Fajfer:2015ycq,Becirevic:2015asa,
		Varzielas:2015iva,Alonso:2015sja,Calibbi:2015kma,Barbieri:2015yvd,
		Becirevic:2016oho,Crivellin:2017zlb,Becirevic:2017jtw,Cai:2017wry}):
	They give tree-level contributions to these processes but contribute,
	for example, to $\Delta F=2$ processes only at the loop level,
	therefore respecting the bounds from other flavour observables.
	Furthermore, since in $R(D^{(*)})$ an $\mathcal{O}(10\%)$ effect
	compared to the tree-level SM is needed, a NP tree-level effect
	is also required. Here, LQ are probably even the most promising solution
	(see for
	example~\cite{Fajfer:2012jt,Bauer:2015knc,Fajfer:2015ycq,Li:2016vvp,
		Becirevic:2016yqi,Sahoo:2016pet,Barbieri:2016las,Chen:2017hir,
		Cai:2017wry,Altmannshofer:2017poe,Dorsner:2017ufx}).  In fact, in
	Ref.~\cite{Crivellin:2017zlb}, a model for a simultaneous explanation
	of $b\to s\mu^+\mu^-$ data together with $R(D^{(*)})$ has been
	proposed which is compatible with the bounds from $B\to
	K^{(*)}\bar{\nu}\nu$, electroweak precision
	data~\cite{Feruglio:2017rjo} and direct LHC
	searches~\cite{Faroughy:2016osc}. Interestingly, LQ also provide a
	natural solution to the anomaly in the magnetic moment of the muon due
	to the possible enhancement by $m_t/m_\mu$ through an internal chirality flipping~\cite{Djouadi:1989md, Chakraverty:2001yg, Cheung:2001ip,
		Bauer:2015knc, ColuccioLeskow:2016dox}.
	
	The model independent fit to $R(K)$ and $R(K^*)$ allows for NP
	contributions to electrons or muons separately, but also to both
	simultaneously~\cite{Altmannshofer:2017yso, DAmico:2017mtc,
		Geng:2017svp, Ciuchini:2017mik, Hiller:2017bzc}. Once the other data
	on $b\to s\mu^+\mu^-$ is included, NP in muons is required but is only
	optional for electrons.  However, the best-fit value suggests a
	simultaneous NP contribution to electrons as
	well~\cite{Capdevila:2017bsm,Altmannshofer:2017yso,Hurth:2017hxg}. It
	is well known that once LQ couple to muons and electrons
	simultaneously, they give rise to lepton flavour violating $B$ decays
	and to $\mu\to e\gamma$~\cite{Varzielas:2015iva} (see
	Fig.~\ref{FeynmanDiagrams}). 
	
	Both $\mu\to e\gamma$ and lepton flavour violating $B$ decays with
	$\mu e$ final states are experimentally very interesting and precise
	upper limits for these processes already exist.  For $\mu \to e
	\gamma$, the current experimental bound, obtained by the MEG
	Collaboration~\cite{TheMEG:2016wtm}, is
	\begin{equation}
	{\rm Br}[\mu \to e \gamma] \leq 4.2 \times 10^{-13}\,,
	\end{equation}
	and MEG II \cite{Baldini:2013ke} at the Paul Scherrer Institute (PSI)
	will significantly improve on this bound in the future. Concerning
	lepton flavour violating $B$ decays with $\mu e$ final states the
	current limits are~\cite{Amhis:2014hma}
	\begin{equation}
	\label{BllEXP}
	\begin{aligned}
	{\rm Br}\left[B^+\to K^+\mu^\pm e^\mp \right]_{\rm exp} &\le& 9.1\times 10^{-8}   \,,\\
	{\rm Br}\left[B\to K^*\mu^\pm e^\mp\right]_{\rm exp} &\le& 1.4\times 10^{-6} \,,\\
	{\rm Br}\left[B_s\to \mu^\pm e^\mp \right]_{\rm exp} &\le& 1.2\times 10^{-8}   \,. 
	\end{aligned}
	\end{equation}
	Also here, LHCb and BELLE II will improve on these bounds in the near
	future.
	
	In this article we examine the interplay between $b\to s\mu^+\mu^-$
	processes, $R(K^{(*)})$, $\mu\to e\gamma$ and $b\to s\mu e$ processes
	in detail considering LQ. For this purpose, we will take into
	account all 10 representations for scalar and vector LQ under
	the SM gauge group. 
	
	The article is structured as follows: In the next section we will fix
	our conventions for the LQ interactions and calculate the
	contributions to $b\to s\ell^+\ell^-$ transitions and $\mu\to
	e\gamma$. We use these results in Sec.~\ref{analysis} to perform a
	phenomenological analysis for the three LQ representations that give a
	good fit to $b\to s \mu^+\mu^-$, considering the most constraining
	processes with electrons and muons in the final state. In
	Sec.~\ref{sec:tau} we briefly comment on $\tau$-$e$ and $\tau$-$\mu$
	transitions before we conclude. The appendix presents the complete
	tree-level matching of the 10 LQ representations for semileptonic $B$
	decays (see also Ref.~\cite{Dorsner:2016wpm,Alonso:2015sja}) and their contributions to all $\ell\to\ell^\prime\gamma$
	processes.
	
	\section{Model and observables}
	\label{sec:observables}
	
	The possible representations of LQ under the SM gauge group
	were first categorized in Ref.~\cite{Buchmuller:1986zs}. There are five
	scalar LQ with the following quantum numbers:
	\begin{equation}
	\begin{aligned}
	Q\left( {\Phi _1^{}} \right)&:&\;\left( {3,1, - \frac{2}{3}} \right)\,,\\
	Q\left( {\tilde \Phi _1^{}} \right)&:&\;\left( {3,1, - \frac{8}{3}} \right)\,,\\
	Q\left( {{\Phi _2}} \right)&:&\;\left( {\bar 3,2, - \frac{7}{3}} \right)\,,\\
	Q\left( {{{\tilde \Phi }_2}} \right)&:&\;\left( {\bar 3,2, - \frac{1}{3}} \right)\,,\\
	Q\left( {{\Phi _3}} \right)&:&\;\left( {3,3, - \frac{2}{3}}\right)\ \ 
	\end{aligned}
	\end{equation}
	under the SM gauge group $SU(3)_C\times SU(2)_L\times U(1)_Y$, respectively.
	These new scalars couple to SM fermions in the following way:
	\begin{equation}
	\begin{aligned}
	\mathcal{L}_{{\rm{scalar}}}^{LQ} =& \left( {\lambda
		_{fi}^{1R}\overline {u_f^c} {\ell _i} + \lambda _{fi}^{1L}\overline{Q_f^c} \I{\tau _2}{L_i}} \right)\Phi _1^\dag + \tilde \lambda _{fi}^1\overline {d_f^c} {\ell _i}\tilde \Phi _1^\dag + \tilde \lambda _{fi}^2{\overline{d _f}}\tilde \Phi _2^\dag{L_i}\\
	& + \left( {\lambda _{fi}^{2RL}{\overline{{u}_f}}{L_i} + \lambda _{fi}^{2LR}\overline {{Q_f}} \I{\tau _2}{\ell _i}} \right)\Phi _2^\dag+ \lambda _{fi}^3\overline {Q_f^c} \I{\tau _2}{\left( {\tau
			\cdot \Phi _3^{}} \right)^\dag }{L_i} + {\rm{h.}}{\rm{c.}}\,.
	\end{aligned}
	\end{equation}
	Here we assumed that lepton number and/or baryon number is conserved.
	This forbids couplings of LQ to two quarks (which are in
	principle allowed by gauge invariance) and ensures the stability of
	the proton.
	
	Concerning vector LQ there are also five representations
	under the SM gauge group with charges
	\begin{equation}
	\begin{aligned}
	Q\left( {V_1^\mu } \right)&:&\;\left( {\bar 3,1, - \frac{4}{3}} \,\right)\,,\\
	Q\left( {\tilde V_1^\mu } \right)&:&\;\left( {\bar 3,1,-\frac{{10}}{3}} \right)\,,\\
	Q\left( {V_2^\mu } \right)&:&\;\left( {3,2, - \frac{5}{3}}\, \right)\,,\\
	Q\left( {\tilde V_2^\mu } \right)&:&\;\left( {3,2,+\frac{1}{3}}\, \right)\,,\\
	Q\left( {V_3^\mu } \right)&:&\left( {3,3, +\frac{4}{3}}\, \right)\,.
	\end{aligned}
	\end{equation}
	These new massive vectors couple to fermions via
	\begin{equation}
	\begin{aligned}
	\mathcal{L}_{{\rm{vector}}}^{LQ} =& \left( {\kappa _{fi}^{1L}\overline {Q_f^{}} \gamma_{\mu}{L_i} + \kappa _{fi}^{1R}\overline {{d_f}} \gamma_{\mu}{\ell _i} } \right)V_1^{\mu\dagger} + \tilde \kappa _{fi}^1\overline{u_f} \gamma_{\mu}{\ell_{i}}\tilde V_1^{\mu\dagger}+ \tilde{\kappa}_{fi}^2\overline {u_f^c} \gamma_{\mu}\tilde{V}_2^{\mu\dagger}{L_i} \\
& +\left( { \kappa _{fi}^{2RL}\overline {d_f^c}\gamma_{\mu}{L_i} +  \kappa _{fi}^{2LR}\overline {Q_f^c} \gamma_{\mu}{\ell _i}} \right)V_2^{\mu\dagger} + \kappa _{fi}^3\overline {{Q_f}}\gamma_{\mu}{\left( {\tau  \cdot V_3^{\mu}} \right)}{L_i} + {\rm{h.}}{\rm{c.}}\,.
	\end{aligned}
	\end{equation}
	Again, we assume the conservation of lepton and baryon number. Even
	though massive vector bosons are not renormalizable without a Higgs
	mechanism, we will not specify the scalar sector. As we will see
	later, this is not necessary for our purpose because the new Higgs
	sector can be decoupled. We point out that this only works because
	$\ell\to\ell^\prime\gamma$ is finite in unitary gauge.  
	
	Let us now turn to the calculation of the most relevant observables,
	$b\to s\mu^+\mu^-$, $b\to s e^+ e^-$, $b\to s\mu e$, and $\mu\to
	e\gamma$.  For reasons explained at the end of this section we set the
	right-handed couplings of LQ to fermions to zero. Furthermore, here we
	give the results solely for the phenomenologically interesting
	representations, $\Phi_3$, $V_1^\mu$ and $V_3^\mu$. Only they give a
	good fit to $b\to s\ell^+\ell^-$ data as they generate left-handed
	currents.  The complete tree-level matching (including right-handed
	couplings) for all LQ representations and all semileptonic $B$ decays
	and $\ell\to\ell^\prime\gamma$ processes can be found in the appendix.
	
	Starting with $b\to s\ell^+\ell^-$ transitions we use the effective
	Hamiltonian
	\begin{align}
	\label{eq:Heff}
	\mathcal{H}_{\text{eff}}^{\ell_f\ell_i} = 
	-\frac{4G_F}{\sqrt{2}}V_{tb}V_{ts}^*\sum_{k}{C_{k}^{fi}O_{k}^{fi}}
	+ {\rm{h.}}{\rm{c.}}
	\end{align}
	restricted to operators with left-handed couplings:
	\begin{equation}
	\label{eq:OpNoprime}
	\begin{aligned}
	O_9^{fi}&=\frac{\alpha}{4\pi}\bar{s}\gamma_{\mu}P_{L}b\, 
	\bar{\ell}_{f}\gamma^\mu\ell_i\,,\\
	O_{10}^{fi}&=\frac{\alpha}{4\pi}\bar{s}\gamma_{\mu}P_{L}b\, 
	\bar{\ell}_{f}\gamma^\mu\gamma_{5}\ell_i\,.\\
	\end{aligned}
	\end{equation}
	The Wilson coefficients $C_{9(10)}^{fi}$ can then be expressed as
	\begin{equation}
	\label{eq:sll}
	\begin{aligned}
	\Phi_3:\; & C_{9}^{fi}=-C_{10}^{fi}     =+\lambda_{3i}^{3} \lambda_{2f}^{3*} \frac{\sqrt{2}}{2G_F V_{tb}V_{ts}^*}\frac{\pi}{\alpha}\frac{1}{M^2}\,, \\
	V_{1}^{\mu}:\; & C_{9}^{fi}=-C_{10}^{fi}=-\kappa_ {2i}^{1L}\kappa_ {3f}^{1L*}\frac{\sqrt{2}}{2G_F V_{tb}V_{ts}^*}\frac{\pi}{\alpha}\frac{1}{M^2}\,, \\
	V_3^{\mu}:\; & C_{9}^{fi}=-C_{10}^{fi}  =-\kappa_ {2i}^{3} \kappa_ {3f}^{3*} \frac{\sqrt{2}}{2G_F V_{tb}V_{ts}^*}\frac{\pi}{\alpha}\frac{1}{M^2}
	\end{aligned}
	\end{equation}
	with the leptoquark mass $M$. The complete results for the Wilson
	coefficients originating for the 10 representations of scalar and
	vector LQ are given in the appendix.
	In order to constrain the Wilson coefficients $C_{9(10)}^{ee,\mu\mu}$
	we use the global fit of Ref.~\cite{Capdevila:2017bsm} to $b\to
	s\ell^+\ell^-$ data.
	
	For the $b\to s\mu e$ transitions we use the results of
	Ref.~\cite{Crivellin:2015era}:
	\begin{align}
		\label{eq:Bsmue}
		{\rm Br}\left[ B_s \to \mu^+ e^- \right] &= \dfrac{\tau_{B_s}
			m_\mu^2 M_{B_s} f_{B_s}^2}{64\pi^3} \alpha^2 G_F^2 \left| V_{tb}
		V_{ts}^*\right|^2 \left(1 - \dfrac{m_\mu^2}{M_{B_s}^2}\right)^2\times\left(\left|
		C_{9}^{\mu e}\right|^2 + \left|
		C_{10}^{\mu e}\right|^2
		\right) \,,\,\quad\,\\
		\label{eq:Kmue}
		{\rm Br}[B\to K^{(*)}\mu^+ e^-] &= 10^{-9} \left(a_{K^{(*)}}
		\left|C_9^{\mu e} \right|^2 + b_{K^{(*)}}\left|C_{10}^{\mu e} \right|^2+ c_{K^{(*)}}\left|C_9^{\mu e}\right|^2 +
		d_{K^{(*)}}\left|C_{10}^{\mu e} \right|^2 \right)\,,
	\end{align}
	with
	\begin{equation}
	\begin{aligned}
	a_{K  } &=15.4 \pm 3.1 \,,  &b_{K  }&=15.7 \pm 3.1 \,,\\
	c_{K  } &= 0           \,,  &d_{K  }&= 0           \,,\\
	a_{K^*} &= 5.6 \pm 1.9 \,,  &b_{K^*}&= 5.6 \pm 1.9 \,,\\
	c_{K^*} &=29.1 \pm 4.9 \,,  &d_{K^*}&=29.1 \pm 4.9 \,.
	\end{aligned}
	\end{equation}
	Note that these results are for $\mu^+e^-$ final states and not for
	the sums $\mu^\pm e^{\mp}=\mu^-e^{+}+\mu^+e^{-}$ that are constrained
	experimentally~\cite{Amhis:2014hma}.
	
	Let us now consider the lepton flavour violating processes $\mu\to
	e\gamma$. Evaluating the loop diagrams depicted in
	Fig.~\ref{FeynmanDiagrams} for the three leptoquark representations in which
	we are interested, we find the branching ratios
	\begin{align}
	\Br[\mu\rightarrow e\gamma]=\tau_\mu\ 
	\frac{\alpha m_{\mu}^{3}}{256\pi^4}\left|C_L^{e\mu}\right|^2
	\label{eq:meg_branching}
	\end{align}
	with
	\begin{equation}
	\label{eq:muegama}
	\begin{aligned}
	\Phi_{3}:\; & C_L^{e\mu}=-N_{c}\dfrac{\lambda_{j1}^{3*}\lambda_{j2}^{3}m_{\mu}}{8M^2}\,,\\
	V_{1}^{\mu}:\; & C_L^{e\mu}=+N_{c}\dfrac{\kappa_{j1}^{1L*}\kappa_{j2}^{1L}m_{\mu}}{6M^2}\,,\\
	V_{3}^{\mu}:\; & C_L^{e\mu}=+N_{c}\dfrac{2\kappa^{3*}_{j1}\kappa^{3}_{j2}m_{\mu}}{M^2}\,.
	\end{aligned}
	\end{equation}
	The complete formula for all leptoquarks is given in the appendix.
	Here we did not follow the approach of Ref.~\cite{Biggio:2016wyy} but
	rather calculated the effect in unitary gauge which gives a UV finite
	result. Note that this is possible since the remaining Higgs sector
	(or additional composite dynamics) can be decoupled such that it does
	not affect $\mu\to e\gamma$.
	
	In general, LQ can also account for the anomalous magnetic moment
	(AMM) of the muon~\cite{Djouadi:1989md, Davidson:1993qk,
		Couture:1995he, Chakraverty:2001yg, Cheung:2001ip, Mahanta:2001yc,
		Queiroz:2014pra, Bauer:2015knc, Das:2016vkr, Biggio:2016wyy,
		ColuccioLeskow:2016dox, Chen:2017hir}.  However, this would require
	chirally enhanced effects which  also enhance $\ell\to\ell'\gamma$
	processes. This enhancement is so large, that $\mu\to e\gamma$ would
	rule out any effect in electrons in $b\to s\ell^+\ell^-$ transitions
	if one accounted for the AMM of the
	muon~\cite{ColuccioLeskow:2016dox}. Therefore, we will assume the
	absence of chiral enhancement in our phenomenological analysis and
	assume that the LQ couple only to left-handed fermions.
	
	In principle also contributions to $\mu\to3e$ arise at the one-loop
	level in LQ models with couplings to $\mu$ and $e$. While the box
	contributions are suppressed by four small LQ-quark-lepton
	couplings (as estimated from the $b\to s\ell^+\ell^-$ anomalies)
	$Z$ penguins are potentially important.  They can lead to
	branching ratios of the order of $10^{-15}$ which is interesting
	in the light of the future expected sensitivity~\cite{Blondel:2013ia}.
	This is due to the contribution of internal top quarks leading to
	an enhancement $m_t^2/m_Z^2$.  However, the same $Z$ penguin also
	generates effects in $\mu\to e$ conversion. In this case also
	tree-level effects can arise, depending on the couplings to the
	first generation of quarks. We postpone a detailed analysis of
	these effects to a forthcoming publication.
	
	LQ also contribute to $b\to s\bar\nu\nu$ and $b\to c\ell\bar\nu$
	transitions. For muons and electrons, these processes do not give
	relevant constraints. However, they are in general important once tau
	leptons are involved and the corresponding formulae are given in the appendix.

	\begin{figure*}[h!]
		\centering	
		\begin{tikzpicture}

		\legline{math3}{1,8.5}{$\mathrm{Br}[B\to K\mu^\pm e^\mp]$ with $\gamma=1/2$}{dashed, very thick}
		\legline{math3}{1,8.0}{$\mathrm{Br}[B\to K\mu^\pm e^\mp]$ with $\gamma=1$}{very thick}
		\legline{math3}{1,7.5}{$\mathrm{Br}[B\to K\mu^\pm e^\mp]$ with $\gamma=2$}{dotted, very thick}
		\legbox{red2  }{1,9}{$b\to s\mu^+\mu^-$ ($2\sigma$)}
		
		\legbox{math1  }{-6,8.5}{$\mathrm{Br}[\mu\to e\gamma] < 4.2\cdot 10^{-13}$ with $\Phi_3$}
		\legbox{math1D }{-6,8}{$\mathrm{Br}[\mu\to e\gamma] < 4.2\cdot 10^{-13}$ with $V_1^\mu$}
		\legbox{math1DD}{-6,7.5}{$\mathrm{Br}[\mu\to e\gamma] < 4.2\cdot 10^{-13}$ with $V_3^\mu$}
		\legbox{red4   }{-6,9}{$b\to s\mu^+\mu^-$ ($1\sigma$)}

		\node at (0,0) {\includegraphics[width=0.8\textwidth]{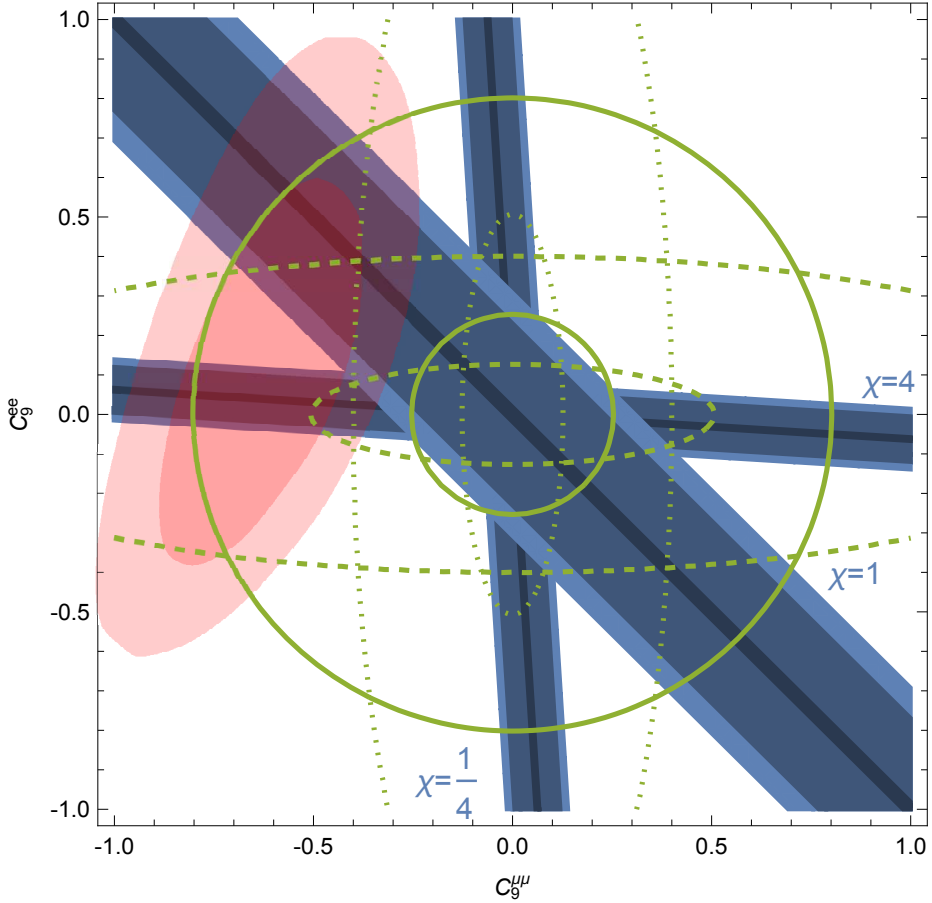}};

		\end{tikzpicture}	
		\caption{
            Regions allowed by MEG (shades of blue) and $b\to
            s\mu^+\mu^-$ (red) in the $C_9^{\mu\mu}$--$C_9^{ee}$
            plane~\cite{Quim} with $C_9=-C_{10}$. The different
            representations are colour-coded in the darkness of the
            different blues: the light-blue region corresponds to
            $\Phi_3$, the medium one to $V_1^\mu$ and the dark blue
            region to $V_3^\mu$. The bands rotated relative to the
            $\chi=1$ region show the situation for $\chi=4$ and
            $\chi=1/4$, respectively.  The green contours represent
            the branching ratio $B\to K\mu^\pm e^\mp$ with $\gamma=1$
            (solid line), $\gamma=1/2$ (dashed) and $\gamma=2$
            (dotted). In each case, the inner line describes
            $\mathrm{Br}[B\to K\mu^\pm e^\mp]=0.2\times10^{-8}$ and
            the outer one $\mathrm{Br}[B\to K\mu^\pm
            e^\mp]=2\times10^{-8}$.  Note that these contours do not
            depend on the specific LQ representation.
		}
		\label{C9C10full}
	\end{figure*}

	\label{analysis}

	\section{Phenomenological analysis}

	As stated above, we focus on the three LQ representations that can
	give a good fit to $b\to s\mu^+\mu^-$ data for the phenomenological
	analysis : $\Phi_3$, $V_1^\mu$, and $V_3^\mu$. In addition, we assume
	that the couplings to right-handed fermions vanish such that all three
	representations give a pure $C_9=-C_{10}$-like contribution.
	Furthermore, we neglect the couplings of the LQ to the first
	generation of quarks. If one takes the deviations from the SM predictions in $b\to c\tau\nu$ processes seriously, the mass scale of the LQ should be around $2$ TeV for perturbative couplings. However, $b\to s\ell^{+}\ell^{-}$ data can also be explained for much heavier LQs (above $10$ TeV) if the couplings are sizable. 
	
	Once the LQ couple to muons and electrons simultaneously, we get
	correlated effects in $\mu\to e\gamma$, $B_s\to \mu e$ and $B\to
	K^{(*)} \mu e$.  Combining \eqref{eq:sll} and \eqref{eq:muegama} with
	\eqref{eq:Kmue} and \eqref{eq:meg_branching} we can express the lepton
	flavour violating branching ratios in terms of the Wilson coefficients
	$C_9^{\mu\mu}$ and $C_9^{ee}$ as
		\begin{align}
		\label{eq:c9c10full}
		\Br[\mu\to e\gamma] &= \tau_\mu\ \frac{\alpha^3 G_F^2
			m_\mu^5}{512\pi^6} |V_{tb}V_{ts}^*|^2 N_{c}^2 \left(\chi C_9^{ee} +
		\frac{C_9^{\mu\mu}}\chi\right)^2 \begin{cases}
		1/16 & \Phi_3\\
		1/9  & V_1^\mu\\
		16    & V_3^\mu
		\end{cases}\,,\\
		\Br[B\to K \mu^\pm e^\mp] &= 10^{-9}\, (a_{K} + b_{K}) \left[
		\left(\frac{C_9^{ee}}\gamma\right)^2 + \left(\gamma\ C_9^{\mu\mu}\right)^2
		\right]\,.
		\end{align}
	Here we defined the ratios $\chi = y_{32}/y_{21}$ and $\gamma =
	y_{21}/y_{22}$, with $y=\lambda$ for scalar LQ and $y=\kappa$ for
	vector LQ.
	
	Note that the constraints from $\mu\to e\gamma$ on the scalar LQ
	triplet is weakest, resulting in the biggest allowed region in
	parameter space and that the effect in $b\to s\mu e$ transitions does
	not depend on the specific representation. Our results are shown in
	Fig.~\ref{C9C10full} for various values of $\chi$ and $\gamma$.
	Interestingly, for real couplings, there is a cancellation in the
	contributions to $\mu\to e\gamma$ if $\sgn C_9^{\mu\mu} = -\sgn
	C_9^{ee}$. This means that if, in the future, the global fit required
	equal signs for $C_9^{\mu\mu}$ and $C_9^{ee}$, a LQ explanation (with
	real couplings) of the anomalies would be ruled out.  Furthermore, the
	predicted rates for $B_s\to\mu e$, $B\to K\mu e$ and $B\to K^*\mu e$
	are within the reach of LHCb and BELLE II. In Fig.~\ref{C9C10full}, we
	only showed $B\to K\mu e$ for which the predicted rate is closest to
	the current experimental limit. For the other processes, we have 
	\begin{equation}
	\begin{aligned}
	{\rm Br}[B\to K^*\mu e]/{\rm Br}[B\to K\mu e] &\approx 2.2\,, \\
	{\rm Br}[B_s\to \mu e]/{\rm Br}[B\to K\mu e] &\approx 0.006 
	\end{aligned}
	\end{equation} 
	in our $C_9=-C_{10}$ setup.
	
	\section{$\tau$-$\mu$ and $\tau$-$e$ transitions}
	\label{sec:tau}
	
	Once one allows for couplings of leptoquarks to tau leptons as well,
	$\tau$-$\mu$ and $\tau$-$e$ transitions are also generated. The
	corresponding processes are experimentally much less constrained than
	$\mu$-$e$ transitions. In fact, the most constraining processes
	involving tau flavours are $B\to K^{(*)}\bar\nu\nu$ which include tau
	neutrinos. In order to generate measurable effects in processes with
	charged tau leptons, the corresponding effect in neutrinos must be
	absent or suppressed. The only single LQ representation which gives a
	good fit to $b\to s\mu^+\mu^-$ data and does not generate effects in
	$b\to s\bar\nu\nu$ is the vector singlet $V_{1}^{\mu}$.  However, this LQ has the
	same tree-level phenomenology as the combination of a scalar singlet
	and a scalar triplet studied in Ref.~\cite{Crivellin:2017zlb}.
	Furthermore, since in the absence of right-handed couplings
	$\tau\to\mu\gamma$ and $\tau\to e\gamma$ are not important, we refer
	the reader to Ref.~\cite{Crivellin:2017zlb} where the interplay
	between $b\to s\tau\mu$, $b\to s\bar\nu\nu$ and  $b\to s\mu^+\mu^-$
	processes is shown.
	
	\section{Conclusions and outlook}\label{conclusions}
	
	In this article we have studied the possibility that LQ contribute to
	$b\to s\mu^+\mu^-$ and $b\to s e^+ e^-$ processes simultaneously in
	order to explain the hints for LFU violation in $R(K)$ and $R(K^*)$,
	generating lepton flavour violation as well. We calculated the
	tree-level matching for semileptonic $B$ decays for all ten (five scalar
	and five vector) LQ representations and their effects at one loop in
	$\ell\to\ell^\prime\gamma$.
	
	In our phenomenological analysis, we considered the three LQ
	representations ($\Phi_3$, $V_1^\mu$ and $V_3^\mu$) giving a good fit
	to $b\to s\ell^+\ell^-$ data. In this setup, we found an interesting
	interplay between $b\to s\ell^+\ell^-$, $\mu\to e\gamma$ and $b\to
	s\mu e$ processes, showing that the current constraints are within the
	same ballpark. The amount of tuning between the electron and the muon coupling of the LQ required by $\mu\to e\gamma$ depends on representation chosen as well as on the ratio $\chi$. In general, the effect of the $\Phi_3$ in $\mu\to e\gamma$ is smallest and therefore less tuning is required than for the other LQs. Interestingly, if forthcoming data requires NP
	contributions to electron and muon channels simultaneously, there are
	also very good prospects of discovering non-zero decay rates for
	processes like $B_s\to\mu e$ or $\mu\to e \gamma$ with measurements in
	the near future. Furthermore, (for real couplings) one could rule out
	a LQ explanation $b\to s\ell^+\ell^-$ if $C_9^{\mu\mu}$ has the same
	sign as $C_9^{ee}$ since this is in conflict with $\mu\to e\gamma$
	bounds.

	\medskip
	
	{\it Acknowledgments} --- {\small 
		The work of A.C. and D.M. is supported by an Ambizione Grant of the
		Swiss National Science Foundation (PZ00P2\_154834). Y.U. is supported
		by the Swiss National Science Foundation (SNF) under contract
		200021\_163466. We are grateful to Bernat Capdevila and Joaquim Matias
		for providing us the fit to $b\to s \ell^+ \ell^-$ for the scenario
		$C_{9}^{\mu\mu}=-C_{10}^{\mu\mu}$ and $C_{9}^{ee}=-C_{10}^{ee}$. We thank Toshihiko Ota and Giovanni Marco Pruna for checking the sign of the Wilson coefficients originating from the tree-level matching. We
		also thank Giovanni Marco Pruna for useful discussions and pointing
		out the consistency of the calculation of $\mu\to e\gamma$ in the
		unitary gauge.}
	
	\section*{Appendix}

In this appendix, we present the tree-level matching for semileptonic
$b\to s$ and $b\to c$ processes and the loop effect in $\ell\to\ell^\prime\gamma$ for all ten
leptoquark representations. Contrary to the results presented in the main article, we keep right-handed couplings.

\renewcommand{\arraystretch}{1.7}
\begin{table*}[t]
	\centering
\scalebox{0.9}{
	\begin{tabular}{|c|c|cc|}
		\hline
		&Representation&$\Gamma_{fi}^{R}$&$\Gamma_{fi}^{L}$\\
		\hline
		$\overline{u_{f}}\ell_{i}$&$\Phi_{2}$&$V_{fj}\lambda^{2LR}_{ji}$&$\lambda_{fi}^{2RL}$\\
		\hline
		$\overline{u_{f}}\nu_{i}$&$\Phi_{2}$&$0$&$\lambda_{fi}^{2RL}$\\
		\hline
		$\overline{d_f}\ell_{i}$&$\Phi_{2}$&$-\lambda_{fi}^{2LR}$&0\\
		&$\tilde{\Phi}_{2}$&0&$\tilde{\lambda}_{fi}^{2}$\\
		\hline
		$\overline{d_{f}}\nu_{i}$&$\tilde{\Phi}_{2}$&$0$&$\tilde{\lambda}_{fi}^{2}$\\
		\hline
		$\overline{u_{f}^{c}}\ell_{i}$&$\Phi_{1}$&$\lambda_{fi}^{1R}$&$V^{*}_{fj}\lambda_{ji}^{1L}$\\
		&$\Phi_{3}$&$0$&$-V^{*}_{fj}\lambda_{ji}^{3}$\\
		\hline
		$\overline{u_{f}^{c}}\nu_{i}$&$\Phi_{3}$&$0$&$\sqrt{2}V_{fj}^{*}\lambda_{jf}^{3}$\\
		\hline
		$\overline{d_{f}^{c}}\ell_{f}$&$\tilde{\Phi}_{1}$&$\tilde{\lambda}^{1}_{fi}$&$0$\\
		&$\Phi_{3}$&$0$&$-\sqrt{2}\lambda_{fi}^{3}$\\		
		\hline
		$\overline{d^{c}_{f}}\nu_{i}$&$\Phi_{1}$&$0$&$-\lambda_{fi}^{1L}$\\
		&$\Phi_{3}$&$0$&$-\lambda_{fi}^{3}$\\
		\hline
	\end{tabular}
	}~~~
	\scalebox{0.9}{
	\begin{tabular}{|r|c|cc|}
		\hline
		&Representation&$\Gamma_{fi}^{VR}$&$\Gamma_{fi}^{VL}$\\
		\hline
		$\overline{u_{f}}\ell_{i}$&$\tilde{V}_{1}^{\mu}$&$\tilde{\kappa}^{1}_{fi}$&0\\
		&$V_{3}^{\mu}$&0&$\sqrt{2}V_{fj}\kappa_{ji}^{3}$\\
		\hline
		$\overline{u_{f}}\nu_i$&$V_{1}^{\mu}$&$0$&$\kappa_{ji}^{1L}V_{jf}$\\
		&$V_{3}^{\mu}$&$0$&$V_{fj}\kappa^{3}_{ji}$\\
		\hline
		$\overline{d_{f}}\ell_i$&$V_{1}^{\mu}$&$\kappa_{fi}^{1R}$&$\kappa_{fi}^{1L}$\\
		&$V_{3}^{\mu}$&$0$&$-\kappa_{fi}^{3}$\\
		\hline
		$\overline{d_{f}}\nu_i$&$V_{3}^{\mu}$&$0$&$\sqrt{2}\kappa_{fi}^{3}$\\
		\hline
		$\overline{u^{c}_{f}}\ell_{i}$&$V_{2}^{\mu}$&$V_{fj}^{*}\kappa^{2LR}_{ji}$&$0$\\
		&$\tilde{V}_{2}^{\mu}$&$0$&$\tilde{\kappa}_{fi}^{2}$\\
		\hline
		$\overline{u^{c}_{f}}\nu_{i}$&$\tilde{V}_{2}^{\mu}$&$0$&$\tilde{\kappa}^{2}_{fi}$\\
		\hline
		$\overline{d^{c}_{f}}\ell_i$&$V_{2}^{\mu}$&$\kappa_{fi}^{2LR}$&$\kappa_{fi}^{2RL}$\\
		\hline
		$\overline{d^{c}_{f}}\nu_{i}$&$V_{2}^{\mu}$&$0$&$\kappa_{fi}^{2RL}$\\
		\hline
	\end{tabular} }
	\caption{Couplings for the different representations of scalar and vector LQ to quarks and leptons.}
	\label{tab:SLQ_coupling_structure}
\end{table*}

\renewcommand{\arraystretch}{1.7}
\begin{table*}[t]
	\centering
\scalebox{0.85}{
	\begin{tabular}{|c|cccccc|}
		\hline
		$b\to s\ell_{i}^{+}\ell_{f}^{-}$&$C_9^{fi}$&$C_{10}^{fi}$&$C_9^{\prime fi}$&$C_{10}^{\prime fi}$&$C_S^{fi}=C_P^{fi}$&$C_S^{\prime fi}=-C_P^{\prime fi}$\\
		\hline
		$\Phi_1$&0&0&0&0&0&0\\
		$\Phi_3$&$2\lambda_{3i}^{3}\lambda_{2f}^{3*}$&$-2\lambda_{3i}^{3}\lambda_{2f}^{3*}$&0&0&0&0\\
		$\Phi_{2}$&$-\lambda_{2i}^{2LR}\lambda_{3f}^{2LR*}$&$-\lambda_{2i}^{2LR}\lambda_{3f}^{2LR*}$&0&0&0&0\\
		$\tilde{\Phi}_{2}$&0&0&$-\tilde{\lambda}_{2i}^{2}\tilde{\lambda }_{3f}^{2*}$&$\tilde{\lambda}_{2i}^{2}\tilde{\lambda}_{3f}^{2*}$&0&0\\
		$\tilde{\Phi}_1$&0&0&$\tilde{\lambda}_{3i}^{1}\tilde{\lambda}_{2f}^{1*}$&$\tilde{\lambda}_{3i}^{1}\tilde{\lambda}_{2f}^{1*}$&0&0\\
		\hline
		$V_{1}^{\mu}$&$-2\kappa_{2i}^{1L}\kappa_{3f}^{1L*}$&$2\kappa_{2i}^{1L}\kappa_{3f}^{1L*}$&$-2\kappa_{2i}^{1R}\kappa_{3f}^{1R*}$&$-2\kappa_{2i}^{1R}\kappa_{3f}^{1R*}$&$4\kappa_{2i}^{1L}\kappa_{3f}^{1R*}$&$4\kappa_{2i}^{1L}\kappa_{3f}^{1R*}$\\
		$V_3^{\mu}$&$-2\kappa_{2i}^{3}\kappa_{3f}^{3*}$&$2\kappa_{2i}^{3}\kappa_{3f}^{3*}$&0&0&0&0\\
		$V_{2}^{\mu}$&$2\kappa_{3i}^{2RL}\kappa_{2f}^{2RL*}$&$2\kappa_{3i}^{2RL}\kappa_{2f}^{2RL*}$&$2\kappa_{3i}^{2LR}\kappa_{2f}^{2LR*}$&$-2\kappa_{3i}^{2LR}\kappa_{2f}^{2LR*}$&$4\kappa_{3i}^{2LR}\kappa_{2f}^{2RL*}$&$4\kappa_{3i}^{2LR}\kappa_{2f}^{2RL*}$\\
		$\tilde{V}_{1}^{\mu}$&0&0&0&0&0&0\\
		$\tilde{V}_{2}^{\mu}$&0&0&0&0&0&0\\
		\hline
	\end{tabular}
}
	\caption{Contribution of the 10 LQ representations to $b\to
		s\ell_{i}^+\ell_{f}^{-}$. Each entry should be multiplied
		by $\frac{\sqrt{2}}{4G_F
			V_{tb}V_{ts}^*}\frac{\pi}{\alpha}\frac{1}{M^2}$.}
	\label{tab:table_b_to_sll}
\end{table*}

In order to simplify the calculation, one can write interactions of LQ
with quarks and leptons completely generic in the following form,
\begin{align*}
\overline{q_f^{(c)}}\left(\Gamma^{R}_{fi}P_{R}\right.&\left.+\Gamma^{L}_{fi}P_{L}\right)\ell_{i}^{(c)}\Phi^{*}_{A}\,,\\
\overline{q_f^{(c)}}\left(\Gamma_{fi}^{VR}\gamma_{\mu}P_{R}\right.&\left.+\Gamma_{fi}^{VL}\gamma_{\mu}P_{L}\right)\ell_{i}^{(c)}V^{\mu *}_{A}\,,
\end{align*}
with
\begin{align*}
\Phi_{A}&\in\{\Phi_{1},\tilde{\Phi}_{1},\Phi_{2},\tilde{\Phi}_{2},\Phi_{3}\}\,,\\
V_{A}^{\mu}&\in\{V_{1}^{\mu},\tilde{V}_{1}^{\mu},V_{2}^{\mu},\tilde{V}_{2}^{\mu},V_{3}^{\mu}\}
\end{align*}
the scalar and vector LQ, respectively. The superscript $(c)$
denotes a possible charge conjugation of the field. The explicit form
of the couplings $\Gamma^{R,L}_{fi}$ and $\Gamma^{VR,VL}_{fi}$ for the
various representations is given in
Table~\ref{tab:SLQ_coupling_structure}. Here, we chose to work in the
down basis, i.e. CKM rotations appear in the couplings once
interactions with left-handed up quarks are present. All other
rotations necessary to go from the interaction to the mass eigenbasis
are unphysical and can be absorbed into a redefinition of the
couplings.

\begin{boldmath}
	\subsection*{$b\to s\ell^+\ell^-$}
\end{boldmath}

For $b\to s\ell^+\ell^-$ transitions we use the effective Hamiltonian in Eq.\eqref{eq:Heff}, also including operators with right-handed couplings,
\begin{align}
\begin{aligned}
O_9^{(\prime)fi}&=\frac{\alpha}{4\pi}\bar{s}\gamma_{\mu}P_{L(R)}b\bar{\ell}_{f}\gamma^\mu\ell_i\,,\\
O_{10}^{(\prime)fi}&=\frac{\alpha}{4\pi}\bar{s}\gamma_{\mu}P_{L(R)}b\bar{\ell}_{f}\gamma^\mu\gamma_{5}\ell_i\,,\\
O_{S}^{(\prime)fi}&=\frac{\alpha}{4\pi}\bar{s}P_{L(R)}b\bar{\ell}_f\ell_i\,,\\
O_{P}^{(\prime)fi}&=\frac{\alpha}{4\pi}\bar{s}P_{L(R)}b\bar{\ell}_f\gamma_{5}\ell_i\,.
\end{aligned}
\end{align}
The Wilson coefficients originating for the ten representations
of scalar and vector LQ are given in Table~\ref{tab:table_b_to_sll}.
Each entry should be understood to be multiplied by a factor
\begin{align}
\frac{\sqrt{2}}{4G_F V_{tb}V_{ts}^*}\frac{\pi}{\alpha}\frac{1}{M^2}\,.
\end{align}

For $i\neq f$, we also get contributions to lepton flavour violating
$B$ decays. 
	\begin{align}
	{\rm Br}\left[ B_s \to \ell^+ \ell^{\prime-} \right] =&\dfrac{\tau_{B_s}
		{\rm Max}[m_\ell^2,m_{\ell^\prime}^2] M_{B_s} f_{B_s}^2}{64\pi^3} \alpha^2 G_F^2 \left| V_{tb}
	V_{ts}^*\right|^2 \left(1 - \dfrac{{\rm Max}[m_\ell^2,m_{\ell^\prime}^2]}{M_{B_s}^2}\right)^2\nonumber\\
	&\times\left(\left|
	C_{9}^{\ell \ell^\prime}-C^{\prime\ell \ell^\prime}_{9}\right|^2 + \left|
	C_{10}^{\ell \ell^\prime}-C^{\prime\ell \ell^\prime}_{10} \right|^2
	\right) \,,\,\quad\,
	\label{bstaumu}\\
	{\rm Br}[B\to K^{(*)}\ell^+\ell^{\prime-}] =&10^{-9} \left(a_{K^{(*)}\ell\ell^\prime}
	\left|C_9^{\ell\ell^\prime} + C_9^{\prime\ell\ell^\prime} \right|^2 +
	b_{K^{(*)}\ell\ell^\prime}\left|C_{10}^{\ell\ell^\prime} +
	C_{10}^{\prime\ell\ell^\prime} \right|^2 \right. \nn &
	 +\left.
	c_{K^{(*)}\ell\ell^\prime}\left|C_9^{\ell\ell^\prime}
	-C_9^{\prime\ell\ell^\prime} \right|^2 +
	d_{K^{(*)}\ell\ell^\prime}\left|C_{10}^{\ell\ell^\prime}
	-C_{10}^{\prime\ell\ell^\prime} \right|^2 \right)\,,
	\label{bkstaumu}
	\end{align}
	with
\vspace{1em}
	\begin{center}
	\scalebox{0.85}{
		\begin{tabular}{|c|c|c|c|c|c|c|c|c|}
			\hline
			$\ell\ell^\prime $ & $a_{K\ell\ell^\prime}$ & $b_{K\ell\ell^\prime}$ &
			$c_{K\ell\ell^\prime}$ & $d_{K\ell\ell^\prime}$ &
			$a_{K^*\ell\ell^\prime}$ & $b_{K^*\ell\ell^\prime}$ &
			$c_{K^*\ell\ell^\prime}$ & $d_{K^*\ell\ell^\prime}$ \\
			\hline
			$\;\tau\mu/\tau e\;$ & $\;9.6 \pm 1.0\;$ & $\;10.0 \pm 1.3\;$ & $0$ & $0$ & $\;3.0 \pm
			0.8\;$ & $\;2.7 \pm 0.7\;$ & $\;16.4 \pm 2.1\;$ & $\;15.4 \pm 1.9\;$
			\\
			$\mu e$ & $15.4 \pm 3.1$ & $15.7 \pm 3.1$ & $0$ & $0$ & $5.6 \pm 1.9$ & $5.6 \pm
			1.9$ & $29.1 \pm 4.9$ & $29.1 \pm 4.9$ \\
			\hline
		\end{tabular}
		}
	\end{center}

\vspace{1em}

Note that the results in~(\ref{bstaumu}) and~(\ref{bkstaumu}) are for
$\ell^-\ell^{\prime+}$ final states and not for the sums
$\ell^\pm\ell^{\prime\mp}=\ell^-\ell^{\prime+}+\ell^+\ell^{\prime-}$
constrained experimentally.

\begin{boldmath}
	\subsection*{$b\rightarrow s\bar{\nu}\nu$}
\end{boldmath}

\renewcommand{\arraystretch}{1.7}
\begin{table}[t]
	\centering
	\vspace{-0.4cm}
	\scalebox{0.9}{
		\begin{tabular}{|c|cc|}
			\hline
			$b\to s\bar{\nu}_{i}\nu_{f}$&$C_L^{fi}$&$C_R^{fi}$\\
			\hline
			$\Phi_1$&$\lambda^{1L}_{3i}\lambda^{1L*}_{2f}$&0\\
			$\Phi_3$&$\lambda^{3}_{3i}\lambda^{3*}_{2f}$&0\\
			$\Phi_{2}$&0&0\\
			$\tilde{\Phi}_{2}$&0&$-\tilde{\lambda}^{2}_{2i}\tilde{\lambda}^{2*}_{3f}$\\
			$\tilde{\Phi}_1$&0&0\\
			\hline
			$V_1^{\mu}$&0&0\\
			$V_3^{\mu}$&$-4\kappa_{2i}^{3}\kappa_{3f}^{3*}$&0\\
			$V_{2}^{\mu}$&0&$2\kappa_{3i}^{2LR}\kappa_{2f}^{LR*}$\\
			$\tilde{V}_{1}^{\mu}$&0&0\\
			$\tilde{V}_{2}^{\mu}$&0&0\\
			\hline
		\end{tabular}
	}
	\caption{Contribution of the various LQ representations to $b\to
		s\bar{\nu}_{i}\nu_{f}$. Each entry should be multiplied by a factor
		$\frac{\sqrt{2}}{4G_F V_{tb}V_{ts}^*}\frac{\pi}{\alpha}\frac{1}{M^2}$.
		\label{tab:table_b_to_snunu}}
\end{table}

Here, we match the Wilson coefficients on the effective Hamiltonian
defined as
\begin{align}
\mathcal{H}_{\text{eff}}^{\nu_{f}\nu_{i}}=-\frac{4G_F}{\sqrt{2}}V_{tb}V_{ts}^{*}\sum_{k}{C_{k}^{fi}O_{k}^{fi}}
+ {\rm{h.}}{\rm{c.}}
\end{align}
with the operators given by
\begin{equation}
\begin{aligned}
O_{L(R)}^{fi}&=\frac{\alpha}{4\pi}\bar{s}\gamma_{\mu}P_{L(R)}b
\, \bar{\nu}_{f}\gamma^{\mu}\left(1-\gamma_5\right)\nu_i\,.\\
\end{aligned}
\end{equation}
The results for the corresponding Wilson coefficients are given in
Table~\ref{tab:table_b_to_snunu} where the overall factor
\begin{align}
\frac{\sqrt{2}}{4G_F V_{tb}V_{ts}^*}\frac{\pi}{\alpha}\frac{1}{M^2}
\end{align}
is omitted.
The ratios between the measurements of $B\to K^{(*)}\bar\nu\nu$ and
the SM
\begin{align}
\mathcal{R}_{K^{(*)}} = 
\frac{\Br[B\to K^{(*)} \bar\nu\nu]}
{\Br[B\to K^{(*)} \bar\nu\nu]_{\rm SM}} \gg 1
\end{align}
are currently much larger than one.\\


\renewcommand{\arraystretch}{1.7}
\begin{table*}[h]
	\centering
	\begin{tabular}{|c|ccccc|}
		\hline
		$b\to c\bar{\nu}_{i}\ell_{f}^{-}$&$C_{VL}^{fi}$&$C_{VR}^{fi}$&$C_{SL}^{fi}$&$C^{fi}_{SR}$&$C_{TL}^{fi}$\\
		\hline
		$\Phi_1$&$-\lambda^{1L}_{3i}V_{2j}\lambda^{1L*}_{jf}$&0&$\lambda^{1L}_{3i}\lambda^{1R*}_{2f}$&0&$-\frac{1}{4}\lambda^{1L}_{3i}\lambda^{1R*}_{2f}$\\
		$\Phi_3$&$\lambda^{3}_{3i}V_{2j}\lambda^{3*}_{jf}$&0&0&0&0\\
		$\Phi_{2}$&0&0&$\lambda^{2RL}_{2i}\lambda^{2LR*}_{3f}$&0&$\frac{1}{4}\lambda^{2RL}_{2i}\lambda^{2LR*}_{3f}$\\
		$\tilde{\Phi}_{2}$&0&0&0&0&0\\
		$\tilde{\Phi}_1$&0&0&0&0&0\\
		\hline
		$V_1^{\mu}$&$-2\kappa_{3f}^{1L*}V_{2j}\kappa_{ji}^{1L}$&0&0&$4\kappa_{3f}^{1R*}V_{2j}\kappa_{ji}^{1L}$&0\\
		$V_3^{\mu}$&$2\kappa_{3f}^{3*}V_{2j}\kappa_{ji}^{3}$&0&0&0&0\\
		$V_{2}^{\mu}$&0&0&0&$4\kappa_{3i}^{2RL}V_{2j}\kappa_{jf}^{2LR*}$&0\\
		$\tilde{V}_{1}^{\mu}$&0&0&0&0&0\\
		$\tilde{V}_{2}^{\mu}$&0&0&0&0&0\\
		\hline
	\end{tabular}
	\caption{Contribution of the various LQ representation to $b\to
		c\bar{\nu}_{i}\ell_{f}^{-}$. Each entry should be multiplied by a
		factor $\frac{-\sqrt{2}}{8G_FV_{cb}}\frac{1}{M^2}$.
		\label{tab:table_b_to_clnu}}
\end{table*}

\renewcommand{\arraystretch}{2.4}
\begin{table*}[h]
	\centering
\scalebox{0.7}{
	\begin{tabular}{|c|c|c|}
		\hline
		$\ell_{i}\to\ell_{f}\gamma$&$C_{L}^{fi}$&$C_{R}^{fi}$\\
		\hline
		$\Phi_{1}$&$\dfrac{\lambda^{1L*}_{jf}\lambda^{1L}_{ji}m_{\ell_{i}}}{24M^2}-\dfrac{\lambda_{jf}^{1R*}V_{jk}^{*}\lambda_{ki}^{1L}m_{u_{j}}\left(7+4\log\left(y_{u_{j}}\right)\right)}{12M^2}$&$\dfrac{\lambda^{1R*}_{jf}\lambda^{1R}_{ji}m_{\ell_{i}}}{24M^2}-\dfrac{V_{jk}\lambda_{kf}^{1L*}\lambda_{ji}^{1R}m_{u_{j}}\left(7+4\log\left(y_{u_{j}}\right)\right)}{12M^2}$\\
		$\tilde{\Phi}_{1}$&$0$&$-\dfrac{\tilde{\lambda}^{1*}_{jf}\tilde{\lambda}^{1}_{ji}m_{\ell_{i}}}{12M^2}$\\
		$\Phi_{2}$&$-\dfrac{\lambda_{jf}^{2RL*}\lambda_{ji}^{2RL}m_{\ell_{i}}}{8M^2}+\dfrac{\lambda_{jf}^{2RL*}V_{jk}\lambda_{ki}^{2LR}m_{u_{j}}\left(1+4\log\left(y_{u_{j}}\right)\right)}{12M^2}$&$-\dfrac{\lambda_{jf}^{2LR*}\lambda_{ji}^{2LR}m_{\ell_{i}}}{8M^2}+\dfrac{V_{jk}^{*}\lambda_{kf}^{2LR*}\lambda_{ji}^{2RL}m_{u_{j}}\left(1+4\log\left(y_{u_{j}}\right)\right)}{12M^2}$\\
		$\tilde{\Phi}_2$&$0$&$0$\\
		$\Phi_{3}$&$-\dfrac{\lambda_{jf}^{3*}\lambda_{ji}^{3}m_{\ell_{i}}}{8M^2}$&$0$\\
		\hline
		$V_{1}^{\mu}$&$\dfrac{\kappa_{jf}^{1L*}\kappa_{ji}^{1L}m_{\ell_{i}}}{6M^2}-\dfrac{\kappa_{jf}^{1R}\kappa_{ji}^{1L}m_{d_{j}}}{3M^2}$&$\dfrac{\kappa_{jf}^{1R*}\kappa_{ji}^{1R}m_{\ell_{i}}}{6M^2}-\dfrac{\kappa_{jf}^{1L*}\kappa_{ji}^{1R}m_{d_{j}}}{M^2}$\\
		$\tilde{V}_{1}^{\mu}$&$0$&$\dfrac{11\tilde{\kappa}^{1*}_{jf}\tilde{\kappa}^{1}_{ji}m_{\ell_{i}}}{12M^2}$\\
		$V_{2}^{\mu}$&$\dfrac{2\kappa_{jf}^{2RL*}\kappa_{ji}^{2RL}m_{\ell_{i}}}{3M^2}-\dfrac{5\kappa_{jf}^{2LR*}\kappa_{ji}^{2RL}m_{d_{j}}}{3M^2}$&$\dfrac{7\kappa_{jf}^{2LR*}\kappa_{ji}^{2LR}m_{\ell_{i}}}{12M^2}-\dfrac{5\kappa_{jf}^{2RL*}\kappa_{ji}^{2LR}m_{d_{j}}}{3M^2}$\\
		$\tilde{V}_{2}^{\mu}$&$-\dfrac{\tilde{\kappa}^{2*}_{jf}\kappa^{2}_{ji}m_{\ell_{i}}}{12M^2}$&$0$\\
		$V_{3}^{\mu}$&$\dfrac{2\kappa^{3*}_{jf}\kappa^{3}_{ji}m_{\ell_{i}}}{M^2}$&$0$\\
		\hline
	\end{tabular}
	}
	\caption{Contribution of the ten LQ representations to
		$\ell_{i}\to\ell_{f}\gamma$ assuming $m_{\ell_{f}}=0$. An
		additional factor $N_{c}$ is understood. For the scalar LQ doublets
		the Wilson coefficients with down-type quarks vanish because of the
		factor $1+3 Q_d$.}
	\label{tab:WC_for_llprimegamma}
\end{table*}

\begin{boldmath}
	\subsection*{$b\rightarrow c\ell\bar{\nu}$}
\end{boldmath}

For completeness, we also consider the charged current effective
Hamiltonian
\begin{align}
\mathcal{H}_{\text{eff}}^{\ell_f\nu_i}=\frac{4G_F}{\sqrt{2}}V_{cb}\sum_{k}{C_{k}^{fi}O_{k}^{fi}}
+ {\rm{h.}}{\rm{c.}}
\end{align}
with
\begin{equation}
\begin{aligned}
O_{VL(R)}^{fi}&=\bar{c}\gamma^{\mu}P_{L(R)}b\, \bar{\ell}_{f}\gamma_{\mu}P_{L}\nu_{i}\,,\\
O_{SL(R)}^{fi}&=\bar{c}P_{L(R)}b\, \bar{\ell}_{f}P_{L}\nu_{i}\,,\\
O_{TL}^{fi}&=\bar{c}\sigma^{\mu\nu}P_{L}b\, \bar{\ell}_{f}\sigma_{\mu\nu}P_{L}\nu_{i}\,.
\end{aligned}
\end{equation}
The Wilson coefficients expressed in terms of the LQ couplings are
given in Table \ref{tab:table_b_to_clnu}, with an overall factor
\begin{align}
\frac{-\sqrt{2}}{8G_FV_{cb}}\frac{1}{M^2}
\end{align}
omitted.

Considering only couplings to muons and electrons, the effects in
$B\to D^{(*)}\ell\nu$ are below the percent level once the constraints
from $b\to s\ell^+\ell^-$ are taken into account and therefore
phenomenologically not relevant.\\

\begin{boldmath}
	\subsection*{$\ell_i\to\ell_f\gamma$}
\end{boldmath}

Here the branching ratios are given by
\begin{align}
\Br[\ell_{i}\rightarrow\ell_{f}\gamma]= \tau_{\ell_i}
\frac{\alpha m_{\ell_{i}}^{3}}{256\pi^4}
\left(\left|C_L^{fi}\right|^2+\left|C_R^{fi}\right|^2\right)\,.
\end{align}
Working with a generic charge $Q$ for the quark propagating in the
loop, we obtain for a vector LQ,
\begin{align}
C_L^{fi}&
=N_{c}\left(
\frac{\Gamma_{jf}^{VL*}\Gamma_{ji}^{VL}m_{\ell_{i}}(5+9Q)}{12M^{2}}
-\frac{\Gamma_{jf}^{VR*}\Gamma_{ji}^{VL}m_{q_{j}   }(1+2Q)}{M^2    }
\right)
\,,
\end{align}
and for a scalar LQ
\begin{align}
C_{L}^{fi}&
=N_{c}\left(-\frac{\Gamma_{jf}^{L*}\Gamma_{ji}^{L}m_{\ell_{i}}(1+3Q)}{24M^{2}}+\frac{\Gamma_{jf}^{L*}\Gamma_{ji}^{R}m_{q_{j}}\left(-1+2Q+2Q\log\left(y_{q_{j}}\right)\right)}{4M^{2}}\right)\,,
\end{align}
where $y_{q_{j}}=m_{q_{j}}^2/M^2$ and $C_R$ is obtained from $C_L$ by exchanging $L$ with $R$. The explicit expressions for
$C_{L}^{fi}$ and $C_{R}^{fi}$ for the various representations after
summing over the $SU(2)$ components are given in
Table~\ref{tab:WC_for_llprimegamma}.
	
\clearpage

\inputencoding{latin2}
\bibliographystyle{JHEP}
\bibliography{BIB}
\inputencoding{utf8}

\end{document}